%% file: ICSE18-NIER.tex
\documentclass[sigconf]{acmart}

\usepackage{booktabs} 
\usepackage{amsmath}

\usepackage{pgfplots}
\usepackage{booktabs}   
\usepackage{subcaption} 
\usepackage{wrapfig}
\usepackage{courier}
\usepackage{tikz}
\usetikzlibrary{arrows}
\usepackage{times}

\usepackage{amssymb}
\usepackage{xspace}
\usepackage{natbib}
\usepackage{amsfonts}
\usepackage{graphicx}
\usepackage{comment}
\usepackage{floatflt}
\usepackage{amsmath}
\usepackage{multicol}
\usepackage{multirow}
\usepackage{enumitem}
\usepackage{color}
\usepackage{listings}
\usepackage{xspace}

\usepackage{algorithm}
\usepackage{algpseudocode}

\usepackage{color}
\usepackage{array}
\usepackage{listings}
\usepackage{xspace}
\usepackage{amsmath}
\usepackage{tikz}
\usepackage{url}
\usepackage{multirow}
\usepackage{stmaryrd}
\usepackage{etoolbox}
\usetikzlibrary{positioning}
\setcopyright{rightsretained}

\renewcommand{\paragraph}[1]{\smallskip\noindent {\bf #1~~}}
\newcommand{\Estimator}{L2S-E}

\newcommand{\figlabel}[1]{\label{fig:#1}}

\newcommand{\tabref}[1]{Table~\ref{tab:#1}}
\newcommand{\tablabel}[1]{\label{tab:#1}}

\begin{document}
\title{Learning to Synthesize}

\author{Yingfei Xiong, Bo Wang, Guirong Fu, Linfei Zang}
\affiliation{
  \institution{Key Laboratory of High Confidence Software Technologies (Peking University), MoE}
  \institution{Institute of Software, EECS, Peking University}
  \city{Beijing}
  \country{China}
  \postcode{100871}
}
\email{{xiongyf,wangbo_15}@pku.edu.cn, {fgr079, zanglf126}@126.com}

\begin{abstract}
  In many scenarios we need to find the most likely program under a
  local context, where the local context can be an incomplete program,
  a partial specification, natural language description, etc. We call
  such problem \emph{program estimation}. In this paper we propose an
  abstract framework, \emph{learning to synthesis}, or \emph{L2S} in
  short, to address this problem. L2S combines four tools to achieve
  this: \emph{syntax} is used to define the search space and search
  steps, \emph{constraints} are used to prune off invalid candidates
  at each search step, \emph{machine-learned models} are used to
  estimate conditional probabilities for the candidates at each search
  step, and \emph{search algorithms} are used to find the best
  possible solution. The main goal of L2S is to lay out the design
  space to motivate the research on program estimation.
  


  We have performed a preliminary evaluation by instantiating
  this framework for synthesizing conditions of an automated program repair (APR) system.
  The training data are from the project itself and related JDK packages.
Compared to ACS, a state-of-the-art condition synthesis system for
program repair, our approach could deal with a larger search space
such that we fixed 4 additional bugs outside the search space of ACS,
and relies only the source
code of the current projects.

\end{abstract}

%
%



\maketitle

\section{Motivation}
\label{sec:introduction}

In many tasks we need to synthesize a program automatically. A typical
category is genetic improvement~\cite{Petke2017Genetic}, where the
system automatically searches for a program within a space to meet a
specific goal, such as performance improvement or bug fixing. Some
problems have clearly-defined specification: when the program meets
the specification, the program is considered as correct. Traditional
program synthesis techniques are mainly designed to deal with such
problems~\cite{Gulwani17}. 
However, many problems do not have such a
correctness specification~\cite{Petke2017Genetic}.
For
example, in test-based program repair~\cite{GenProg} and program by
examples~\cite{Gulwani-IJCAR16}, only a set of tests is available to
validate the correctness of the patched program. Other related fields
are code completion~\cite{Nguyen-ICSE12} and program synthesis from
natural languages~\cite{Raghothaman-ICSE16}, where code is generated
based on a partial program and/or natural language specifications. In
all the above cases, it is not enough to generate a program that
satisfies the specification. For the former two scenarios, existing
studies have revealed repairing only for passing the tests often
results in incorrect patches. In the latter two scenarios, there is
not even a partial specification, and returning an arbitrary
compilable program is definitely not desirable.

A more desirable solution to these cases, as we argue in this paper, is to
find the program that is most-likely to be written under the current
context. More formally, we would like to find a program $prog$ that
maximizes the conditional probability $P(prog \mid context)$, where
$context$ refers to the local context, including a specification, an
incomplete program, and/or natural language description. To
distinguish from the general program synthesis or genetic improvement problems, we call
this problem \emph{program estimation}.

However, it is not easy to solve the program estimation problem.
First, we need to estimate the conditional probability
$P(Prog \mid Context)$. In particular, we need to ensure the estimated
probability is consistent with the partial specification, i.e.,
programs that do not satisfy the specification should have zero
probability. Second, we need to locate the program with the maximum
probability, which is not easy because the space of possible programs
is usually huge.

In this paper we 
propose an abstract framework, \emph{learning to synthesis}, or \emph{L2S} in
short, for program estimation. The main goal of L2S is to lay out the design
space to motivate the research on program estimation. L2S solves the
program estimation problem by combining four tools: syntax,
constraints, machine learned models, and search algorithms. Syntax is
used to define the solution space and to convert the estimation problem into a
search problem. Constraints include the partial specification
and other possible constraints such as type constraints or size
constraints, and are used to prune off the infeasible candidates at
each search step. Machine learned models are used to estimate
conditional probabilities for the candidates at each step, and these
probabilities can be combined to estimate the probability of a generated
program. Finally, the search algorithm is used to solve the search
problem, by using the estimated conditional probabilities as hints.

We have instantiated L2S on a automated program repair (APR) system,
aiming to synthesis correct conditional expressions as patches. We
performed an preliminary evaluation of L2S on two projects from the
Defects4J benchmark, with 133 defects in total. Comparing to
state-of-art APR systems, ACS~\cite{xiong-icse17}, L2S could deal with
a significantly larger search space, leading to 4 additional bugs to
be fixed outside the search space of ACS, and relies only the source
code of the current projects. 

In the rest of the paper, we shall first describe the framework and its implementation in
details, then present our preliminary evaluation, finally discuss
related work and conclude the paper.

\section{Framework}

\subsection{Overview}

We first give an overview of L2S by describing an example of
estimating a conditional expression. Conditional expressions
are a common source of bugs, and existing work~\cite{xiong-icse17,Monperrus:CoRR18}
has shown that correctly estimating conditions could help repair a
significant number of bugs. Now assuming that in the current local
context we have two integer variables, ``hours'' and ``value'', and we
would like to estimate the conditional expression used in the next
``if'' statement. We start by defining the grammar for conditional
expressions.
\[
  \tt
  E \rightarrow E\ ``> 12" \mid  E\ `` > 0" \mid E\ ``+"\ E \mid ``hours" \mid ``value"  \mid \ldots
\]

For demonstration purpose, we only show five rules with a shared left
hand side. This grammar turns the estimation problem into a search
problem by defining a series of search steps to produce an abstract
syntax tree (AST): we start from a tree containing one non-terminal
node $E$, and at each step we choose a non-terminal leaf node and a
rule starting from the non-terminal, and add the right hand side as
children of the leaf node. The search stops when there is no
non-terminal leaf node in the AST. Please note that this conversion
assumes a top-down order to expand the tree, and our framework also
supports other possible orders, e.g., staring from a terminal leaf
node instead of the root
node. 
For clarity, we shall only discuss the top-down order in this
subsection, and will generalize to other orders later.

To reduce the search space, we use constraints to prune off infeasible
choices. At each step, we generate constraints from the
context, the already generated AST, the chosen node to expand, and the
chosen production rule. Then we put the constraints
into a constraint solver to check their satisfiability. If
unsatisfiable, this expansion is an infeasible choice. For example, a
common class of constraints is the type constraints. Since we are
generating a Boolean expression, by using the type constraints we know
that only the first two production rules are feasible to expand the root node.

To distinguish the feasible choices at each step, we use machine
learning to calculate the conditional probabilities of each choice.
The conditional probability has the form
$P(Rule | Context, Prog, Node)$, where $Context$ represents the
context for the program generation, $Prog$ represents the currently
generated AST, $Node$ represents the non-terminal node chosen to be
expanded, and $Rule$ represents the choice of a rule starting from the symbol of
$Node$. Please note we do not need to compare choices across different
non-terminal nodes, as all non-terminal nodes will be expanded in the
end.

L2S does not enforce any concrete machine learning methods, and the
user could choose the methods that fit best to the problem. Furthermore,
different machine learning methods could be specified for different
non-terminals for best results. To train the models, L2S requires a
training set includes pairs of programs and their contexts, and parses
the programs to produce the training set at each non-terminal, similar
to PHOG~\cite{Bielik-ICML16}. For each non-terminal node in the parsed
AST, the chosen production rule is a positive instance, and all other
production rules starting from the same non-terminal are negative
instances.

Given the conditional probability of the choice at each step, the
probability of the whole program is their product. Please note that a
program can be generated in different ways by choosing different
non-terminals to expand at search steps, but ideally the probabilities
calculated from different orders will be the same. A detailed proof
will be presented later.

Now we can estimate the probability of each program, we need to solve
the search problem to select the most-likely program. 
Please note that we cannot simply select the most probable choice at
each step, because local optimal does not necessarily lead to global
optimal. For example, let us assume at the root node the learned model
estimates the following probabilities. 
\[
\begin{array}{ll}
  \tt  E \rightarrow E\ ``> 12"  &  0.3 \\
  \tt E \rightarrow E\ ``> 0"  &  0.6\\                                 
\end{array}
\]

Please note other options have been pruned off by type constraints.
When choosing $\tt E\ ``> 0"$ that has the highest probability, the
learned model estimates the following probabilities for the newly
added $\tt E$ node.
\[
\begin{array}{ll}
  \tt  E \rightarrow ``hours"  &  0.1 \\
  \tt  E \rightarrow ``value"  &  0.2 \\
  \tt  E \rightarrow E\ ``+"\ E  &  0.05 \\
\end{array}
\]

Combining the two, the most likely expression is \texttt{value > 0} which has a probability
of 0.12. However, if we choose the other option $\tt E\ ``> 12"$ at
the first step, the learned model predicts the following probabilities
for the newly added $\tt E$ node.
\[
\begin{array}{ll}
  \tt  E \rightarrow ``hours"  &  0.8 \\
  \tt  E \rightarrow ``value"  &  0.1 \\
  \tt  E \rightarrow E\ ``+"\ E  &  0.05 \\
\end{array}
\]

Thus, the combination \texttt{hours > 12} actually has a higher
probability of 0.24. If we select only the best candidate at each
step, we would not be able to produce this expression.



At each step we need to make two decisions. First select a
non-terminal node, and then select a rule to expand it. For the
former, L2S uses a policy that does not depend on the
machine-learned models. We require the policy not to depend on the learned
models so that we can use this policy in preparing the training set.
For the latter, L2S uses a search algorithm to find a set of proper
choices at all steps to maximize the probability of the synthesized
program. L2S does not enforce any particular policy or search
algorithm and the user could choose those suitable to the problem. For
example, a policy could be used in our example is to expand the
non-terminals from left to right, and a search algorithm could be used
is beam search~\cite{beamsearch}. 
The beam search is a greedy algorithm that keeps at most k ASTs that have
the highest probability. At each step, the algorithm constructs a new set of
ASTs by expanding the next 
non-terminal with the k best rules in each AST, and keeps k new ASTs with the highest
probability. 
In the
above example, if we set $k$ to 2, we shall get \texttt{hours > 12}.

In the rest of this section we discuss several key issues in L2S. 

\subsection{Different Expansion Orders}
\label{sec:syntax}

In the previous subsection we have seen how to use syntax to convert
the program estimation problem by expanding the AST in a top-down
order. However, top-down order may not result in the best performance.
For example, when estimating a conditional expression, it is often
easier to first predict which variable should be used in the expression,
and then predict which operation should be applied on the variables
based on our experience. Thus, we need to generalize the framework to
different orders.

To support upward expansion, we introduce two annotations to each node
$N$ in an AST. $N^{D}$ indicates that $N$ can be expanded downward by
adding a subtree where $N$ is the root, as
in the top-down order expansion we have seen. $N^{U}$ indicates that
$N$ can be expanded upward by adding a subtree where $N$ is a leaf.
Similarly, $N$ indicates that the node does need further expansion
while $N^{UD}$ indicates that expansions are needed in both directions.

With the additional annotations, we can generalize grammar rules into
rewriting rules. The original grammar rules perform downward
expansions, and we can generalize them into 
\emph{top-down rewriting rules} as follows.
\[
  \tt
  E_{0}^{D} \Rightarrow {E_{0}} \rightarrow E^{D}\ ``> 12" \mid
  {E_{0}} \rightarrow E^{D}\ `` > 0" \mid {E_{0}} \rightarrow ``hours" \mid \ldots
\]
The symbol before $\Rightarrow$ (the left hand side) indicates that a node with that symbol
can be replaced with the right hand side, where $\tt E_{0} \rightarrow
E^{D}\ ``> 12"$ indicates a tree with $\tt {E}$ as
parent and $\tt E^{D}$ and $\tt ``> 12"$ as children. The subscript
$\tt 0$ in $\tt E_{0}$ indicates that this node will replace the
original node matched by the left hand side.

We
can also generate \emph{bottom-up rules} to expand nodes upward.
\[\begin{array}{lll}
  \tt
  E_{0}^{U}  &\Rightarrow & \tt {E^{U}} \rightarrow {E}_{0}\ ``> 12" \mid {E^{U}} \rightarrow {E_{0}}\ ``> 0" \\
        & \tt \mid &\tt {E^{U}} \rightarrow {E_{0}}\ `` + "\ E^{D} \mid {E^{U}} \rightarrow E^{D}\ `` + "\ {E_{0}} \\
  \tt ``hours"_{0}^{U}  &\Rightarrow & \tt E^{U} \rightarrow {``hours"_{0}} \\
    \tt ``value"_{0}^{U}  &\Rightarrow & \tt E^{U} \rightarrow {``value"_{0}} \\
    \tt E_{0}^{U} &\Rightarrow& \tt E_{0}
\end{array}\]
For each production rule that can generate a node with symbol $N$, we
create a bottom-up rewriting rule starting from $N^{U}$. The rewriting rule
adds a subtree where $N$ is a leaf node. The last rule is generated
for the root symbol to finish the expansion.

With both top-down rules and bottom-up rules, we can expand a node in
both directions. However, to enable bottom-up expansion, we need to
start from a node that is not the root. So we can further introduce
\emph{creation rules}.
\begin{alignat}{3} 
    &\tt\Rightarrow E^{D} & \textit{//Creating a root node.} \label{eq:3} \\
    &\tt\Rightarrow ``value"^{U} & \textit{//Creating a leaf node.} \label{eq:4}\\
    &\tt\Rightarrow E^{UD}& \textit{//Creating a node in the middle.} \label{eq:5}
\end{alignat}
The creation rules do not have a left hand side, indicating it can be
applied to create a tree.
By choosing a proper annotation, we can
create a root node, a leaf node, or a node in the middle.

To ensure that we construct one AST, we require only one application
of creation rules. We call this ``one-tree'' expansion. L2S also
supports ``multi-tree'' expansion, where several ASTs could be
constructed independently and then connected together.
The basic idea is to introduce \emph{connecting rules} such as
the following ones.
\[\begin{array}{lll}
\tt    (E_{0}^{U}, E_{1}^{U})\Rightarrow E^{U} \rightarrow E_{0}\ ``+"\ E_{1}  & \textit{//Horizontal connection.} \\
\tt    (E_{0}^{D}, E_{1}^{U})\Rightarrow E_{0} \rightarrow E^{D}\ ``+"\ E_{1}  & \textit{//Vertical connection.} \\
  \end{array}\]
The rules matches two nodes, each in a different AST tree, and
connect them using the trees in the right hand sides. The notes with
subscripts on the right hand side will replace the nodes with the same
subscripts on the left hand side. Please note that we can only connect
nodes in two ASTs but not two nodes in
the same AST, as ill-formed tree will be generated.


Though technically all the above rewriting rules can be used, too many choices would unnecessarily complicate the
search space. The user should select a set of rules that are most
suitable to the target problem. In particular, we concern
unambiguous set of rules. Given an AST tree and a set of rewriting
rules $R$, in general multiple sequences of rule applications may generate the
tree. The rule set $R$ is \emph{unambiguous} if and only if for any
node $n$ in an arbitrary AST, when $n$ is expanded upward/downward in
one sequence, it will be expanded upward/downward by the same rule in
all sequences. In other words, each AST can be constructed by only one
set of rewriting rule applications, but orders of the applications can
be different. For examples, the top-down rules plus rule (\ref{eq:3})
form an unambiguous set. The top-down rules, bottom-up rules
excluding $\tt E_{0}^{U} \Rightarrow {E^{U}} \rightarrow {E_{0}}\ `` +
"\ E^{D}$, and rule (\ref{eq:4}) form another unambiguous set. 
The property of unambiguousness is important as it allows the
calculation for the
probability of an AST, as shown later. 

Since we generalize production rules into rewriting rules, the other
parts of the framework should also be adjusted to the new search
space, i.e., the machine-learned models calculate the
probabilities of rewriting rule choices for expanding each node, while
the search algorithm uses the rewriting rules to define search steps.

\subsection{Constraint Generation}
\label{sec:constraints}
L2S introduces a structural way of generating the constraints based on
syntax. Given an AST, variables are generated from AST nodes while
constraints are generated from the rewriting rules for constructing
the AST and/or the context. To demonstrate, let us consider the AST of
the expression \texttt{hours > 12}, where two rewriting rules are used
to generate the AST.
\begin{align}
  \tt E_{1}^{D} \Rightarrow E_{1} \rightarrow E_{2}^{D}\ `` > 12" \label{eq:1}\\
  \tt E_{2}^{D} \Rightarrow E_{2} \rightarrow ``hours" \label{eq:2}
\end{align}

We start from the type constraints. Type constraints can be generated
using the standard type inference algorithms.
First, each node $n$ in the tree generates a type variable $T[n]$,
which is an enumeration of types. Then the following two constraints
are generated from the two rewriting rules.
\begin{align*}
  \tt \ T[E_{2}]=Int\ & \textit{//generated from rule (\ref{eq:1})}\\
  \tt T[E_{2}] = T[``hours"] \ & \textit{//generated from rule (\ref{eq:2})}
\end{align*}
And the context gives us the following constraint.
\[
\tt  T[``hours"] = Int,\ T[E_{1}] = Boolean
\]
In this case the constraints are satisfiable, so there is no type
error. However, if we try to expand $E_{1}$ with $E \rightarrow
``hours"$, the constraint will be unsatisfiable and we know this
expansion is infeasible.

Similarly, in the scenarios of test-based program repair and
programming by example, we can generate variables from the nodes to
represent their values in test executions, and generate constraints
based on program semantics and the test cases. In this way, if a
partial program could not satisfy a test case, we could know its
infeasibility early.

Another interesting type of constraints is the size constraints. In
many usage scenarios, we would like to limit the size of the generated
ASTs to avoid searching a too large space. Let us assume the size is
defined as the number of nodes in an AST, we can first statically
calculate the lower bound for expanding each symbol using the
following rules. For simplicity, we assume no connecting rules.
\[\begin{array}{lll}
  size_{s}(symbol^{A})&=& min(\{size_{t}(tree) \mid \textrm{exists rule } symbol^{A} \Rightarrow tree \}) \\
  size_{s}(symbol)&=& 1 \\
  size_{n}(node) &= & size_{s}(s_{node}) \ \textrm{where
                 }s_{node}\textrm{ is the
                 symbol of }node\\
  size_{t}(tree)&=& sum(\{size_{n}(n) \mid \textrm{for each node }n\textrm{ in
               }tree \})
  \end{array}
  \]
Then given an AST $t$, we require $size_{t}(t)$ is smaller
than the limit.

Please note that L2S requires that the generated constraints are
solvable by a constraint solver. In particular, the constraint solver
should support incremental solving, such that we can efficiently check multiple
candidate rules.

\subsection{Machine Learning}
\label{sec:machine-learning}
After generalizing the production rules into rewriting rules, the role
of machine-learned models is to discriminate the set of rewriting rule
sharing the same left hand side. In other words, the model estimates
the conditional probability $P(Rule\mid Context, Prog, Node)$ where
$Rule$ represents the choice of a rewriting rule starting from the
symbol of $Node$ while the other three random variables have the same
meaning as before.

The user specifies the machine
learning algorithm and the functions to extract features from the
partial AST and the context, and provides a training set consisting of
programs and their contexts. For each program in the training set, L2S
parses the program, finds the sequence of rule applications to
generate the program based on the policy of choosing symbols to
expand, and produces a set of training
set using the feature extract functions. 

\subsection{Probability of a Program}
\label{sec:search-algorithm}
In this subsection we discuss how to calculate the probability of a program. Here
we assume a program can be uniquely parsed and do not distinguish
between a program and its AST. 
When the rewriting rules are
unambiguous, the probability of the program, $P(Prog \mid Context)$, is
the product of the conditional probabilities\footnote{To facilitate the presentation, we 
assume that a creation rule application also expands a pseudo node.},
$P(Rule\mid Context, Prog, Node)$, of each rule choice made along any
generation process.

Now we show why the calculation is correct. Since $Context$ is always
available as a condition, we ignore $Context$ in the discussion.
First, since the grammar
is unambiguous, each AST $prog$ is decided by the choices of node to expand
$n_{1}, n_{2}, \ldots, n_{m}$ and the choice of rule at each node
$r_{1}, r_{2}, \ldots, r_{m}$. Thus,
$P(prog)=P(n_{1}, \ldots, n_{m}, r_{1},\ldots,r_{m})$. Now let us
assume the nodes are expanded in the order $(i_{1}, i_{2}, \ldots,
i_{m})$, and the AST after each node expansion are $(prog_{1}, \ldots,
prog_{m})$ where $prog_{m}=prog$ . We have
\[\begin{array}{lll}
  P(prog)&=&P(n_{1}, \ldots, n_{m},
             r_{1},\ldots,r_{m})\\
         &=&P(n_{i_{1}})P(r_{i_{1}}\mid n_{i_{1}})P(n_{i_{2}}
             \mid n_{i_{1}}, r_{i_{1}}) \ldots P(r_{i_{m}}\mid \\
    &&\quad n_{i_{1}},\ldots,n_{i_{m}},r_{i_{1}}\ldots,r_{i_{m-1}}).
\end{array}\]
      We notice
the probabilities $P(n_{i_{1}})$, $P(n_{i_{2}}
\mid n_{i_{1}}, r_{i_{1}})$, \ldots,$P(n_{i_{m}}
\mid n_{i_{1}}, \ldots,
  n_{i_{m-1}}, r_{i_{1}},\ldots,r_{i_{m-1}})$ must all be 100\% because the
corresponding nodes, $n_{i_{1}}\ldots n_{i_{m}}$, have been generated and must be expanded in the
process. We also notice that
\[
  P(prog_{k})=P(n_{i_{1}}, \ldots,
  n_{i_{k}}, r_{i_{1}},\ldots,r_{i_{k}}),
\]
then we have
\[P(prog)=P(r_{i_{1}}\mid n_{i_{1}})P(r_{i_{2}}\mid prog_{1},
  n_{i_{2}})\ldots P(r_{i_{m}}\mid prog_{m-1}, n_{i_{m}}),\] which is
the product from the probability of each rule choice along the order.
Since the order is arbitrarily chosen, any order could lead to the
same probability.

Please note that we rely on machine-learned models to estimate the
conditional probability. However, many discriminative machine-learned
models do not calculate probabilities. In addition, when the rule set is
ambiguous, we cannot calculate the probability in this way. In
these cases, the user needs to provide a fitness function to tell L2S
that how to score AST trees based on the outputs of the
machine-learned models.

\input{impl}

\input{eval}

\section{Related Work}
\label{sec:related-work}

Several studies try to build prediction model for code, either based on
probabilistic CFG~\cite{Bielik-ICML16} or based on graph models~\cite{Nguyen-ICSE15}. Different from us, these
approaches focus on predicting the next element in the model rather
than a whole AST.


Several studies focus on generating API usage code
snippet~\cite{Weimer-ICSE12,Raghothaman-ICSE16}. Generally, these
approaches first try to extract abstract patterns of API usage from
code, find the pattern most relevant to the query, and then map the
pattern back into code. Different from these approaches, L2S is guided
by syntax and uses
machine learning to select a rule at each step.

In the domain of program synthesis, recently several attempts~\cite{Gulwani-APLAS17,Singh-SNAPL17,Murali-CoRR17} have
been made to also use machine learning to generate programs under the
guidance of a syntax. Compared with those approaches, which are
designed to solve a particular problem, L2S is designed as a general framework
for laying out a design space of program estimation. This results in two
main differences. First, each of these approaches covers only part of
the issues discussed in this paper. For example, all these approaches
use top-down expansion and do not consider other orders. Second, some
of these approaches rely properties of their target problem and
cannot be easily generalized to other problems. For example, in
\citet{Gulwani-APLAS17}'s work, each choice at a search step should be
able to transform the input for use in the next step, which does not
hold in general.


\section{Conclusion}
\label{sec:conclusion}

In this paper we have seen a framework, learning to synthesize, for
program estimation. From practical perspective, the framework allows
users to design an algorithm to solve a concrete program estimation
problem by instantiating the components in the framework. From
academic perspective, L2S lays out a design space of program
estimation, which would hopefully facilitate and inspire new research
in this area.

\bibliographystyle{ACM-Reference-Format}
\bibliography{PLDE-bib/plde} 

\end{document}

%% file: impl.tex
\section{Estimating Conditions}
To understand the potential of this framework, we instantiate
L2S framework for conditional expression synthesis. The instantiated
synthesizer is called \emph{\Estimator}. As studied by
Victor et al. \cite{Monperrus:CoRR18}, in Defects4J
\cite{just2014defects4j}, a widely used real-world Java bug
dataset, 42.78\% bugs are related to conditional blocks, which are the
most prevalent category. Estimating conditional expressions has multiple potential usage
scenarios, such as bug fixing, bug detecting, and code completion. In
this paper we mainly focus on bug fixing, in which case we need to
estimate a new condition to replace the localized buggy condition. 

Our instantiation perform cross-project training to complete a missing
conditional expression in a project. The input to our implementation
is the Java source code of a project, where one conditional expression
is missing. Our implementation first uses the source code of the
target project to train a set of prediction models, and then based on
these models to search for a conditional expression at the target
location. For simplicity, we consider only conditional expressions
without logic operators (``\texttt{\&\&}'', ``\texttt{||}'', and
``\texttt{!}''). Conditional expressions such as \texttt{a \&\& !b}
will be treated as two expressions \texttt{a} and \texttt{b}.

\subsection{Syntax}

\begin{figure}[http]
	\centering
          $\begin{array}{lcl}
 \tt           E & \rightarrow&\tt  \tau_{1}(V_1,\dots, V_{k_{1}})\mid\ \ldots\ \mid\tau_{n}(V_1,\dots, V_{k_{n}})\\
\tt	V &\rightarrow&\tt var_{1}\mid\ldots\mid var_{m}
	\end{array}$
	\caption{\label{fig:general-syn} The Meta Syntax of Conditional Expression }
\end{figure}

\begin{figure}[http]
	\centering
	$\begin{array}{lcl}
\tt            E & \rightarrow&\tt  V\ ``>12" \mid V\ ``>0" \mid V\ ``+" ~V~ ``>0" \mid \ldots\\
\tt	V &\rightarrow&\tt ``hours" \mid ``value" \mid \ldots
	\end{array}$
	\caption{\label{fig:concrete-syn} A Concrete Syntax of Conditional Expression }
\end{figure}

Figure~\ref{fig:general-syn} shows the meta grammar for our condition
synthesis. The meta grammar will be instantiated using the data in the
training set and in the local context. Figure~\ref{fig:general-syn}
shows an instantiated grammar. More concretely, the non-terminal $E$
expands to an expression with variables represented by non-terminal
$V$. In the right hand side, each $\tau_{i}$ represents a conditional
expression used in the training dataset. The non-terminal $V$ expands
to a variable in the local context of the conditional expression.
Compared with the running example in the previous section, this
grammar flattens the hierarchy of $E$ and uses a two-level expansion:
the first level expands $E$ to an expression that contains only
non-terminal $V$, and the second level expands each $V$ to a variable.
This goal of this flattening is to simplify the construction of
machine-learning models.

In our instantiation we use bottom-up order to complete the
conditional expression. More concretely, Figure~\ref{fig:general-syn}
shows the rewriting rule we derive from the syntax to synthesize the
conditional expressions. We first predict the left most variable,
which is a leaf of an AST. Then we upward predict an expression as the
parent of the variable. At last we predict the remaining variable
downward. Finally we get an conditional expression.
Figure~\ref{fig:general-syn} further depicts two examples that we
synthesize using this order. In this figure, the circled numbers
indicate the order of synthesis while the arrows show the structure of
the AST. 

\begin{figure}[http]
	\centering
        $\begin{array}{lcl}
\tt           &\Rightarrow&\tt V^{U} \rightarrow var_{1} \mid \cdots
                         \mid V^{U} \rightarrow var_{m}\\
\tt           V^{U}_{1} &\Rightarrow&\tt E \rightarrow
                               \tau_{1}(V_1,V_{2}^{D}\cdots,
                               V^{D}_{k_{1}})\mid\ \cdots\ \\
  \tt         &\mid&\tt E \rightarrow\tau_{n}(V_1,V^{D}_{2},\cdots, V^{D}_{k_{n}})\\
  \tt         V^{D}_{1} &\Rightarrow& \tt
           V_{1} \rightarrow var_{1}\mid\ldots\mid var_{m}
	\end{array}$
	\caption{\label{fig:rewriting} The Meta Rewriting Rules }
\end{figure}

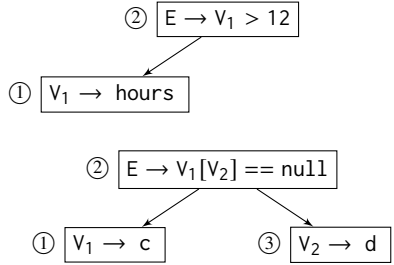
\begin{figure}[http]
	\begin{tikzpicture}
	
	\tikzset{vertex/.style = {shape=rectangle, draw}}
	\tikzset{edge/.style = {->,> = latex'}}
	
	\node[label=left:\textcircled{\footnotesize{2}}] [vertex] (a) at
        (0,1) {$\tt E\rightarrow V_1 > 12$};
	\node[label=left:\textcircled{\footnotesize{1}}] [vertex] (c) at
        (-1.5,0) {$\tt V_1\rightarrow~hours~$};

	\draw[edge] (a) to (c);
	
	\node[label=left:\textcircled{\footnotesize{2}}] [vertex] (d) at
        (0,-1) {$\tt E\rightarrow \tt V_1[V_2] == null$};
	\node[label=left:\textcircled{\footnotesize{1}}] [vertex] (e) at
        (-1.5,-2) {$\tt V_1\rightarrow~c~$};
	\node[label=left:\textcircled{\footnotesize{3}}] [vertex] (f) at
        (1.5,-2) {$\tt V_2\rightarrow~d~$};
	
	\draw[edge] (d) to (e);
	\draw[edge] (d) to (f);
	\end{tikzpicture}
	
	\caption{AST Examples of the Syntax}\figlabel{ast-example}
\end{figure}

\subsection{Constrains}
In our instantiation we use type constraints to filter out
infeasible options. Our implementation reads the type declarations and
variable declarations
from the source file of the target project, and generates type
constraints based on the variable and types used in the expressions.
An example of the type constraints has already been given in
Section~\ref{sec:constraints}. Please note that here we do not need
size constraint since the size of the conditional expressions have
already been confined by the conditional expressions in the training
set. We also do not use value constraint as in this scenario, value
constraints seldom reduces the size of the search space.

\subsection{Machine-Learned Models}
There are three set of rewriting rules in Figure~\ref{fig:rewriting},
each sharing the same left hand side. To assign probabilities to the
rules in each set, we need to build three prediction models, one for
each set. For the ease of presentation, we share call the three sets
as creation rules, expression rules, and variable rules, and call the
three models as the creation model, the expression model, and the
variable model. 

In the creation model and the variable model, we need to predict the
probabilities of variables. In the expression model, we need to
predicate the probabilities of expressions. While the expressions are
extracted from the training set, the variables are extracted from the
local context and may not be available in the training set. As a
result, we need to use different machine learning models to deal with
them. For expressions, we consider each expression as a class and
train a multiclass classifier to predict the probability of each
class. For variables, we consider map each variable to a feature vector,
and train a binary classifier to predict the probability of this
variable.



\subsubsection{Feature Engineering}
Our implementation uses four set of features, as detailed below.
\begin{description}
\item [Context Features] Context features are extracted from the
  context of condition to be synthesized.
\item [Variable Features] Variable features are extracted from a
  variable to represent different aspects of the variable.
\item [Expression Features] Expression features are extracted from an
  expression to represent different aspects of the expression.
\item [Position Feature] This feature is mainly used by variable model
  to indicate the index of the variable being considered.
\end{description}

The features used by the three models are the combinations of the
above four sets. The creation model uses the context features to
represent the context and the variable features to represent the current
chosen variable. The expression model uses the context features to
represent the context and the variable features to represent the
variable that has been chosen. The variable model uses all the four
sets of features, where the context features are used to represent to
the context, the variable features are used to represent the
previous variable in the expression, the expression features are used
to represent the chosen expression, and the position feature is used
to indicate the position of the variable being considered.

In many of the features, we need to encode names, such as variable
name, method name, type name, etc. One important property we would
like to achieve is to let similar names have similar encodings. For
example, ``length'' is often abbreviated into ``len'', and we would
like to give a similar encoding to them. To achieve this, we extract a
bag containing 2-gram of characters from the names, and then uses a
vector to represent the bag. The vector has $n\times n$ dimensions,
where $n$ is the number of characters that can be used in names. Each
dimension in this vector represents the occurrence count of a
corresponding 2-gram. For example, a variable ``len'' contains two
2-grams, ``le'' and ``en'', and then we have a vector where the
dimension of ``le'' and ``en'' are 1 and all other dimensions are 0.
Finally, to reduce dimensions, we apply principle component
analysis~\cite{Deutsch2002Principle} on our training set to map the
name vectors to vectors with no more than 20 dimensions.


Below we give details about the four sets of features. 


\paragraph{Context features} Context features 
include 
class information, method information and code structure information.

Class information indicates features of the containing Java class,
includes basic information such as package name, class name, field
names (all field names encoded as one vector) and field types, and
complexity features such as inheritance hierarchy depth, class length
and method number.

Method information includes the features of the enclosing method, such
as its name, return type, modifiers, lines of method body, parameters
and etc.

Code structure information captures the syntax structure at a certain
line of a java program, including token vectors~\cite{WangLT16} before
and after conditional expression.

\paragraph{Variable features}
Variables features mainly consist of four aspects: naming information,
type information, definition information, documentation information,
definition information and usage information.

Naming information captures the information in the variable name. First, we encode
variable names as vectors using the method mentioned above. Second, we
argument this vector with the length of name, the last word in the
name, etc.


Type information includes features related to the type of the
variable. 
First, the type name is encoded as a vector as mentioned above. 
Second, we argument the vector with a classification of the types. We
classify the type names into integer type (\texttt{short, int, lang}),
float point type (\texttt{float, double}), array type (arrays),
collection type (subclasses of \texttt{List}, \texttt{Set} and
\texttt{Map}), string related type (\texttt{char,  String,
  StringBuffer}) and other type, and use an integer to represent each
type. 
Third, we argument other features, such as a Boolean feature to
indicate whether the name of variable is contained in its type, e.g.,
\texttt{map} is contained in \texttt{Hashmap}.

Definition information is extracted from the declaration of a
variable, including whether it is changeable (\texttt{final}), whether
it is static (\texttt{static}), whether it is loop index/iterator and
its initialization value if available, the distance between the
variable definition and the conditional expression. 

For some well documented projects, their document contains plenty of
useful information. We analyze whether this variable is mentioned in
specific patterns in the Java document, such as throwing an exception.


Definition information concerns how the value in the variable is
defined. We extract at most $k$ places where the variable would be
defined (a variable may be defined in multiple places following
different paths), each with a feature representing the type of the
expression producing the value.

Usages information concerns how the variable is used, including
features such as how many times the variable is used in other if
conditional expressions in the same method, the same class, or the
whole project. 


\paragraph{Expression  features} 
Expression features are extracted from an expression (possibly
containing nonterminals), such as how many variables it contains, the
expected type of the first $k$ variables, invoked method name, used
comparable operators, arithmetic operators and numbers. 

\paragraph{Position feature} The last category includes only one
feature indicating the index of the variable to be expanded. 


\subsubsection{Model Training}
In our implementation we utilize the Gradient Boosting Tree algorithm
to train the three
models. 
Tree-based algorithms are suitable for our task, because they can
handle unbalanced data easily and do not need complexly preprocess on
training set. We select XGBoost \cite{chen2016xgboost}, which a widely
used implementation of the algorithm.



\subsection{Search Algorithms}
To solve the search problem, we need to decide the policy for
selecting non-terminal and the search algorithm for selecting the
rule. For non-terminals, we simple expand the non-terminals from
left to right. For rules, we use the beam search algorithm. After the
creation rules, we keep the top 5 results, and after expanding the
expression, we keep the top 200 results.


When applying our approach to repair conditional expressions, we also
use anti-patterns~\cite{tananti} to disable expressions that easily
lead to incorrect but plausible repairs. In our implementation one
anti-pattern is used: \texttt{obj != null}. This pattern is disabled
because such expressions usually evaluated as \texttt{true}, which can
easily generate plausible conditions. 




%% file: eval.tex
\section{Preliminary Evaluation}
\subsection{Experimental Setup}
We have performed a preliminary evaluation of \Estimator.
Our evaluation is based on two projects in from the
Defects4J~\cite{just2014defects4j} benchmark. Defects4J is a benchmark
of real-world defects in large software projects. The two projects we
used are apache-commons-math and joda-time, both containing a rich set
of conditional expressions. The two subjects have 133
bugs in total. \tabref{subjects} manifests the details of the
subjects.

The evaluation consists of two experiments. In the first experiments,
we took two versions of the two projects, math-12 and time-11. The two
 versions are selected because they are the largest programs in all
 trained subjects in the second experiment. 
For
each version, we randomly
selected 10\% conditions as testing set, and the rest of the conditional
expressions as training set. Then we trained \Estimator
using the train set, and used \Estimator to estimate the conditions
in the test set.

In the second experiment, we replaced the condition synthesizer
component in ACS~\cite{xiong-icse17} with \Estimator to repair the defects in the
benchmark. ACS is a program repair system that focuses on repairing
incorrect conditions. It either inserts new if statements for dealing
with boundary cases, or arguments existing if conditions with
additional Boolean expressions. In both repair patterns, conditional
expressions need to be synthesized and ACS originally uses a
synthesizers that query the whole GitHub repository database to
predicate the most-likely expression. In our experiments we replace
the synthesizer with \Estimator. Furthermore, while in
the second case ACS argument an existing if condition, our modified
ACS directly replaces the original condition for more readable results.

\begin{table}[ht]
	\centering
	\caption{Statistics of the Used Subjects of Defects4J \tablabel{subjects}}
	\begin{tabular}{ccccc}
		\hline
		Project & KLoc & Test Cases & Defects & ID\\
		\hline
		Commons-Math & 104 & 3602 & 106 & Math\\
		Joda-Time & 81 & 4130 & 27 & Time\\
		\hline
		Total & 185 & 7732 & 133 &-\\
		\hline
	\end{tabular}\\
\end{table}



To save training time, instead of training an estimator for each bug,
we use the same trained estimator for a set of bugs that are close in time.
More concretely, we train an estimator using the first version during
a calendar year, and fix the later version based on the estimator.
\tabref{usde-versions} gives the details of the training
versions and the corresponding repaired versions. 

\begin{table}
	\caption{Training Versions \tablabel{usde-versions}}
	\begin{tabular}{ll}
		\hline
		Training & Repaired\\
		\hline
		Math12 & Math1-Math12\\
		Math37 & Math13-Math37\\
		Math59 & Math38-Math59 \\
		Math75 & Math60-Math75 \\
		Math94 & Math76-Math94 \\
		Math102 & Math95-Math102 \\
		Math104 & Math103-Math104 \\	
		Math106 & Math105-Math106 \\
		Time11 & Time1-Time11 \\
		Time17 & Time11-Time17 \\
		Time24 & Time18-Time24 \\
		Time27 & Time25-Time27 \\			
		\hline
	\end{tabular}
\end{table}

All the experiments ran on a personal computer with Intel Core i7-6700
3.4GHz CPU, 16G memory, Ubuntu 16.04LTS and JDK 1.7. We used 30
minutes as the timeout for each defect, same as the experiments of ACS~\cite{xiong-icse17}.

\subsection{Experimental Results}

\begin{table}
  \caption{Predict Results \tablabel{predres}}
\begin{tabular}{cccc}
  \hline
  Project & Top 1 Precision  & Top 10 Precision  & Top 50 Precision \\
  \hline
  Math12 & 37.8\% & 62.2\% & 69.9\%\\
  Time11 & 48.9\% & 71.2\% & 75.5\%\%\\
  \hline
  Average & 43.5\% & 66.7\% & 72.7\%\\
  \hline
\end{tabular}
\end{table}

\tabref{predres} presents the experiment of predicting conditions within a project.
The results show that the average precision can be 43.5\% for top one,
66.7\% for top ten, and 72.7\% for top fifty. 
The result is consistent with existing studies~\cite{Hindle2012} that the code is
repetitive and predictable.

\begin{table}[t]
	\centering
	\caption{Overall Comparison with Existing Techniques \\(Correct~/~Incorrect)\tablabel{overall}}
	\begin{tabular}{ccccc}
		\hline
		Technique & Commons-Math & Joda-Time & Total\\
		\hline
		\Estimator & 9~/~12 & 2~/~4 & 11~/~16\\
		ACS & 12~/~4 & 1~/~0 & 13~/~4\\
		Nopol & 1~/~20 & 0~/~1 & 1~/~21\\
		\hline
	\end{tabular}\\
\end{table}

\tabref{overall} presents the overall results of the experiment on defect
repair. We compare our result with ACS~\cite{xiong-icse17} and Nopol~\cite{Martinez2017}, which are state-of-art techniques focusing on if-statement repair. 

\tabref{overall} presents the overall comparison results in terms of
number of correctly and incorrectly fixed defects of the two subjects.
Recording to \tabref{overall}, we can find that ACS repaired the most
defects correctly with the least incorrectly repaired defects. L2S correctly
repaired two defects less than ACS while generated more wrong patches.
Nopol repaired the least defects with the most wrong patches. Please
note that, while ACS requires the whole Github repository as backend,
L2S and Nopol requires only the source code of the current project. 

\tabref{detail} presents the repair results in detail. L2S has 4
patches that the others cannot repair. 
Further studying the four patches we found that all these patches are
outside the search space of ACS. In order to minimize the wrong
patches, ACS uses a very small search space without method calls or
arithmetic operations. On the other hand, the search space of
\Estimator is much larger including all predicates appeared in
conditional expressions.

\begin{table}[http]
	\centering
	\caption{Detailed Analysis of Patches\tablabel{detail}}
	\begin{tabular}{ccccc}
		\hline
		Bug ID & \Estimator & ACS & Nopol\\
		\hline
		Math2 & Incorrect & -- & -- \\
		Math3 & Correct & Correct & -- \\
		Math4 & Correct & Correct & -- \\
		Math5 & Correct & Correct & -- \\
		Math25 & Correct & Correct & -- \\
		Math28 & Incorrect & Incorrect & -- \\
		Math32 & \textbf{Correct} & -- & Incorrect \\
		Math33 & \textbf{Correct} & -- & Incorrect \\
		Math35 & Correct & Correct & -- \\
		Math40 & -- & -- & Incorrect \\
		Math41 & Incorrect & -- & -- \\
		Math42 & -- & -- & Incorrect \\
		Math46 & Incorrect & -- & -- \\
		Math48 & Incorrect & -- & -- \\
		Math49 & -- & -- & Incorrect \\
		Math50 & Incorrect & -- & Correct \\
		Math57 & -- & -- & Incorrect \\
		Math58 & -- & -- & Incorrect \\
		Math61 & Correct & Correct & -- \\
		Math63 & \textbf{Correct} & -- & -- \\
		Math64 & Incorrect & -- & -- \\
		Math69 & -- & -- & Incorrect \\
		Math71 & Incorrect & -- & Incorrect \\
		Math78 & -- & -- & Incorrect \\
		Math80 & -- & -- & Incorrect \\
		Math81 & -- & Incorrect & Incorrect \\
		Math82 & -- & \textbf{Correct} & Incorrect \\
		Math85 & Incorrect & \textbf{Correct} & Incorrect \\
		Math87 & -- & -- & Incorrect \\
		Math88 & -- & -- & Incorrect \\
		Math89 & -- & \textbf{Correct} & -- \\
		Math90 & -- & \textbf{Correct} & -- \\
		Math97 & -- & Incorrect & Incorrect \\
		Math99 & Incorrect & \textbf{Correct} & -- \\
		Math104 & -- & -- & Incorrect \\
		Math105 & -- & -- & Incorrect \\
		Time1 & Incorrect & -- & -- \\
		Time11 & -- & -- & Incorrect \\
		Time15 & Correct & Correct & -- \\
		Time18 & Incorrect & -- & -- \\
		Time19 & \textbf{Correct} & -- & -- \\
		\hline
	\end{tabular}\\
	\parbox{\columnwidth}{\footnotesize{``Correct'' means the patch is success. ``Incorrect'' means the patch is a wrong patch. ``--''
	 means there is no patch generated. The bold ``Correct'' word means it is only fixed by one approach.}}
\end{table}
